\documentclass[10pt]{article}
\usepackage{amssymb,latexsym,amsmath,enumerate,verbatim,amsfonts,amsthm}
\usepackage{cite}
\usepackage[ansinew]{inputenc}
\usepackage{graphicx,color,float,caption}
\usepackage{pgfplots}
\usepackage{tikz}
\usetikzlibrary{positioning, shapes.geometric}
\newtheorem{theorem}{Theorem}
\newtheorem{corollary}[theorem]{Corollary}
\newtheorem{proposition}[theorem]{Proposition}

\newtheorem{remark}{Remark}[section]
\begin{document}
\title{Vacuum stability conditions  of the general two-Higgs-doublet potential}
\author{Yisheng Song\thanks{Corresponding author. School of Mathematical Sciences, Chongqing Normal University, Chongqing 401331 P.R. China, Email: yisheng.song@cqnu.edu.cn}}
       
\date{}
\maketitle
\abstract{  In this paper, we present the novel analytical expressions for the bounded-from-below or the vacuum stability conditions  of scalar potential for a general  two-Higgs-doublet model  by using the concepts of co-positivity and the gauge orbit spaces.  More precisely, several 
	 sufficient conditions and necessary conditions are established for the vacuum stability of the general 2HDM potential, respectively. We also give an equivalent condition of the vacuum stability of the general 2HDM potential in theory, and then, apply it to derive the  analytical necessary conditions of the general 2HDM potential. Meanwhile, the positive semi-definiteness is proved for a class of 4th-order 2-dimensional complex tensor.}

\section{Introduction}

  In 1973,  the fist two-Higgs-doublet model (for short, 2HDM) is presented by Lee \cite{L1973,L1974}.  Subsequently,  Weinberg \cite{SW} proposed a general multi-Higgs potential model. Since then,  the stability of the scalar Higgs potential is an important problem with the Standard Model (for short, SM) at high-energies.  The bounded-from-below (for short, BFB) or the vacuum stability of SM  is very noticeable in particle physics community.  One of the simplest extensions of the SM Higgs sector is the 2HDM \cite{L1973,DM}.
    It is well-known that the most general Higgs potential for a 2HDM with Higgs doublets $\Phi_1$ and $\Phi_2$ can be written \cite{WW,PW,BFLRS,IS2015} 
    \begin{equation}\label{eq:V}\aligned V_H(\Phi_1,\Phi_2)=&\mu_{11}\Phi_1^*\Phi_1+\mu_{22}\Phi_2^*\Phi_2-(\mu_{12}\Phi_1^*\Phi_2+\mu_{12}^*\Phi_2^*\Phi_1)\\ &+\lambda_1(\Phi_1^*\Phi_1)^2+\lambda_2(\Phi_2^*\Phi_2)^2\\&+\lambda_3(\Phi_1^*\Phi_1)(\Phi_2^*\Phi_2)+\lambda_4(\Phi_1^*\Phi_2)(\Phi_2^*\Phi_1)\\
    	&+\frac{\lambda_5}2(\Phi_1^*\Phi_2)^2+\frac{\lambda_5^*}2(\Phi_2^*\Phi_1)^2\\&+(\Phi_1^*\Phi_1)(\lambda_6\Phi_1^*\Phi_2+\lambda_6^*\Phi_2^*\Phi_1)\\
    	&+(\Phi_2^*\Phi_2)(\lambda_7\Phi_1^*\Phi_2+\lambda_7^*\Phi_2^*\Phi_1),
    	\endaligned\end{equation}
    where $\Phi^*$ is Hermitian conjugate of $\Phi$. The parameters $\mu_{11}$, $\mu_{22}$ and $\lambda_i\ (i=1,2,3,4)$ are real, $\mu_{12}$ and $\lambda_i\ (i=5,6,7)$ are complex.  

    The Higgs potential of such a 2HDM \eqref{eq:V} may be said \cite{BS,BFLRS}
    \begin{equation}\label{eq:2}V_H(\Phi_1,\Phi_2)=\sum_{a,b=1}^{2}\mu_{ab}\Phi_a^*\Phi_b+\sum_{i,j,k,l=1}^{2}t_{ijkl}(\Phi_i^*\Phi_j)(\Phi_k^*\Phi_l),\end{equation}
    where, by definition, \begin{equation}\label{eq:3}t_{ijkl}=t_{klij},\  t_{ijkl}=t_{jilk}^*, \ \mu_{ab}=\mu_{ba}^*.\end{equation}
    The quartic part,  \begin{equation}\label{eq:4}V_4(\Phi_1,\Phi_2)=\sum\limits_{i,j,k,l=1}^{2}t_{ijkl}(\Phi_i^*\Phi_j)(\Phi_k^*\Phi_l),\end{equation}
     gives a 4th-order 2-dimensional complex tensor $\mathcal{T}=(t_{ijkl})$:
    \begin{equation}\label{eq:T}
    	\aligned
    	t_{1111}=&\lambda_{1}, \ t_{2222}=\lambda_{2}, \\
    	t_{1122}=&t_{2211}=\dfrac12\lambda_3, \ t_{1221}=t_{2112}=\dfrac12\lambda_4, \\
    	t_{1212}=&\dfrac12\lambda_5, \ t_{2121}=\dfrac12\lambda_5^*, \\
    	t_{1112}=&t_{1211}=\dfrac12\lambda_6, \ t_{1121}=\ t_{2111}=\dfrac12\lambda_6^*, \\
    	t_{1222}=&t_{2212}=\dfrac12\lambda_7, \ t_{2122}=\ t_{2221}=\dfrac12\lambda_7^*.
    	\endaligned
    \end{equation}

The stability of the 2HDM potential requires that there is no direction in field space along which the potential tends to minus infinity, i.e.,  it is the BFB.  In general,  the quartic part of the scalar potential, $V_4$, is non-negative for arbitrarily large values of the component fields, but the quadratic part of the scalar potential, $V_2$, can take negative values for at least some values of the fields\cite{BFLRS}.
Considering only the quartic part $V_4$, the condition for stability (the BFB) of the scalar potential in the 2HDM is equivalent to the co-positivity or semi-positive definiteness of the tensor $\mathcal{T}=(t_{ijkl})$ given by the Higgs quartic coupling $\lambda_i$, i.e. $V_4(\Phi_1,\Phi_2)\geq 0$. When $\lambda_6 = \lambda_7 = 0$,  the vacuum stability conditions of 2HDM potential\cite{DM,GK2005,K,NS,KKO,ERS} are the following:
$$\lambda_1>0, \lambda_2>0, \lambda_3+2\sqrt{\lambda_1\lambda_2}\geq0, \lambda_3+\lambda_4-|\lambda_5|+2\sqrt{\lambda_1\lambda_2}\geq0.$$
In 2011, Battye et al \cite{BBP} employed the method using Sylvester's criterion and Lagrange multipliers  for the general CP-violating 2HDM.
 In 2016,  Kannike\cite{K2012,K2016,K2021} presented the vacuum stability conditions of the  scalar potential of two Higgs doublets in the 2HDM with explicit CP conservation.  Chauhan\cite{GC} derived analytical necessary and sufficient conditions for the vacuum stability of the left-right symmetric model, and gave the sufficient  conditions for successful symmetry breaking. Recently, Song\cite{S2022} showed the analytical sufficient and necessary conditions of the co-positivity  of the tensor $\mathcal{T}=(t_{ijkl})$ with the real nembers $\lambda_i\ (i=5,6,7)$,  and moreover,  the vacuum stability conditions of scalar potential for the 2HMD with explicit CP conservation was obtained. Bahl et.al.\cite{BCCIW} gave  the analytical sufficient conditions of the vacuum stability for the  2HMD  potential with CP conservation and CP violation, respectively,  where the vacuum stability condition for the  2HMD  potential with CP violation depends on the Lagrange multiplier $\zeta$. 
For more details about  the BFB or the vacuum stability conditions of the 2HDM potential, see Refs. \cite{BFIS,I2007, K,N2020,GH,GOO} for  2HDM with CP conservation; Refs.\cite{I2007,IS2015} for the most general 2HDM; Refs. \cite{BFLRS,I2007,BCCIW} for 2HDM with CP conservation and CP violation; Ref. \cite{MMNN} for 2HDM handled numerically  and others references that are no cited here.

 In this paper, we provide three new analytical sufficient conditions for the bounded-from-below or the vacuum stability of scalar potential for a general  2HDM  by using the co-positivity of 4th-order 2-dimensional symmetric real tensor, which is different from the ones of Bahl et.al.\cite{BCCIW}.  For example,  $$\aligned	& \lambda_1=\lambda_2=1, \lambda_3=-1, \lambda_4=2,  \lambda_5=-1, \lambda_6= \lambda_7=-i.\\
 &\mbox{Obviously, } \textbf{Re}\lambda_6=\textbf{Re}\lambda_7=\textbf{Im}\lambda_5=0, \lambda_3+2\sqrt{\lambda_1\lambda_2}=-1+2>0,\\ &\lambda_3+\lambda_4+\textbf{Re}\lambda_5+2\sqrt{\lambda_1\lambda_2}=-1+2-1+2>0,\\
 &\Delta=0,\ -2\sqrt{\lambda_1\lambda_2}\leq \lambda_3+\lambda_4-\textbf{Re}\lambda_5 \leq6\sqrt{\lambda_1\lambda_2}.
 \endaligned $$
 That is, these parameters meet the conditions (\textbf{II}) and  (\textbf{III}), which means $V_4(\Phi_1,\Phi_2)\geq0$. It is obvious that the conditions $(\textbf{IV}')$ and  (\textbf{III}) are fulfiled also. Here, the conditions (\textbf{II}), (\textbf{III}) and $(\textbf{IV}')$ will be derived in Section 3.  However, they can't satisfy the condition Eq. (5.20) of Bahl et.al. [22], i.e.
 $$\aligned	& 3\sqrt{\lambda_1\lambda_2}-(\lambda_3+|\lambda_4|+|\lambda_5|)=3-(-1+2+1)>0,\\ &\sqrt{\lambda_1\lambda_2}+\lambda_3-(|\lambda_4|+|\lambda_5|+4\left|\lambda_6\sqrt[4]{\dfrac{\lambda_2}{\lambda_1}}\right|)=1-1-(2+1+4)<0,\\
 &\sqrt{\lambda_1\lambda_2}+\lambda_3-(|\lambda_4|+|\lambda_5|+4\left|\lambda_7\sqrt[4]{\dfrac{\lambda_1}{\lambda_2}}\right|)=1-1-(2+1+4)<0.
 \endaligned $$  A sufficient and necessary condition  of the vacuum stability of the general 2HDM potential is given in theory, which contains the vacuum stability condition of the general 2HDM potential with  $\mathbb{Z}_2$ symmetry as a special case. Then, we apply this conclusion to derive the   analytical  necessary conditions of the vacuum stability of a general  2HDM  scalar potential. Meanwhile, the analytical sufficient conditions and necessary conditions are obtained for the semi-positive definiteness of  a class of 4th-order 2-dimensional complex tensor.

 \section{Co-positivity criteria}

 The co-positivity of a matrix $\textbf{M}=(\mu_{ij})$ has been applied to test the vacuum stability  of the 2HDM in Refs. \cite{K2016,K2012,K2021,GC}. It is kown-well that a $2\times 2$ symmetric real matrix $\textbf{M}=(\mu_{ij})$ is co-positive, i.e., \mbox for all non-negative vectors $\textbf{x}=(x_1,x_2)^\top\in \mathbb{R}^2$, the quadratic form  $$\textbf{x}^\top \textbf{M}\textbf{x}=\mu_{11}x_1^2+2\mu_{12}x_1x_2+\mu_{22}x_2^2\geq0,$$ if and only if \cite{ACE,H1983,N1992}   \begin{equation}\label{eq:6}\mu_{11}\geq0,\mu_{22}\geq0 \mbox{ and } \mu_{12}+\sqrt{\mu_{11}\mu_{22}}\geq0.\end{equation}

The co-positivity of a symmetric real tensor has been used to the SM in literature to obtain vacuum stability conditions in Refs. \cite{S2022,K2016,LS2022,SQ2021,SL2022,S2021}. A $4$th-order $n$-dimensional symmetric real tensor $\mathcal{T}=(t_{ijkl}) $ is co-positive \cite{Q2005,Q2013, QSZ,SQ2015,QCC2018,QL2017} if for all non-negative vectors ${\bf x}=(x_1,x_2,\cdots x_n)^\top\in \mathbb{R}^n$,  the quartic form
$$\mathcal{T}{\bf x}^4=\sum_{i,j,k,l=1}^nt_{ijkl}x_{i}x_jx_k x_l\geq0.$$
Let $t_{1111}=\alpha_0>0$ and $t_{2222}=\alpha_4>0.$  Recently, Song and Li \cite{SL2022} presented the analytical expressions of co-positivity of a 4th-order 2-dimensional symmetric $\mathcal{T}$ with the help of the update version (\cite{QSZ}) of Ulrich and Watson’s result \cite{UW}. Let  $\mathcal{T}=(t_{ijkl}) $ be a  $4$th-order $2$-dimensional symmetric real tensor with its entires
$$t_{1111}=\alpha_0, t_{2222}=\alpha_4, t_{1112}= \frac{1}{4}\alpha_1, t_{1122}= \frac{1}{6}\alpha_2,
	  t_{1222}=\displaystyle\frac{1}{4}\alpha_3.
$$
Then the quartic form \begin{equation}\label{eq:06}
	\mathcal{T}{\bf x}^4=\alpha_0 x^4_1+\alpha_1x_1^3x_2+\alpha_2x_1^2x_2^2+\alpha_3x_1x_2^3+\alpha_4x_2^4\geq0
\end{equation} for all $x_1\geq0, x_2\geq0$ if and only if
	$$\aligned	(1) & \Delta\leq0, \alpha_1\sqrt{\alpha_4}+\alpha_3\sqrt{\alpha_0}>0;\\
		(2) & \alpha_1\geq0, \alpha_3\geq0,  2\sqrt{\alpha_0\alpha_4}+\alpha_2\geq0;\\
	(3) & \Delta\geq0, |\alpha_1\sqrt{\alpha_4}-\alpha_3\sqrt{\alpha_0}|\leq4\sqrt{\alpha_0\alpha_2\alpha_4+2\alpha_0\alpha_4\sqrt{\alpha_0\alpha_4}},\\
	& \quad(i) -2\sqrt{\alpha_0\alpha_4}\leq \alpha_2 \leq6\sqrt{\alpha_0\alpha_4},\\
		&\quad(ii)  \alpha_2>6\sqrt{\alpha_0\alpha_4}\\
	& \quad\alpha_1\sqrt{\alpha_4}+\alpha_3\sqrt{\alpha_0} \geq -4\sqrt{\alpha_0\alpha_2\alpha_4-2\alpha_0\alpha_4\sqrt{\alpha_0\alpha_4}},
	\endaligned$$
	where $\Delta=4(12\alpha_0\alpha_4-3\alpha_1\alpha_3+\alpha_2^{2})^{3}-(72\alpha_0\alpha_2\alpha_4+9\alpha_1\alpha_2\alpha_3-2\alpha_2^{3}-27\alpha_0\alpha_3^{2}-27\alpha_1^{2}\alpha_4)^{2}$.\\
	
	Song and Qi \cite[Theorem 3.7]{SQ2021} gave a stronger sufficient  condition for the co-positivity of a symmetric real tensor $\mathcal{T}=(t_{ijkl}) $.  That is, $\mathcal{T}{\bf x}^4\geq0$ for all $x_1\geq0, x_2\geq0$ if
\begin{equation}\label{eq:07}\aligned	& \beta=\alpha_1+4\sqrt[4]{\alpha_0^3\alpha_4}\geq0, \gamma=\alpha_3+4\sqrt[4]{\alpha_0\alpha_4^3}\geq0,\\
& \alpha_2-6\sqrt{\alpha_0\alpha_4}+2\sqrt{\beta\gamma}\geq0.
\endaligned\end{equation}

Song\cite{S2022} obtained  an analytical sufficient and necessary condition for the co-positivity of a symmetric real tensor $\mathcal{T}(\rho,\theta)=(t_{ijkl}(\rho,\theta))$ with two parameters $\rho\in[0,1]$ and $\theta\in [0,2\pi]$, \begin{equation}\label{eq:7}\aligned
t_{1111}=&\Lambda_{1}, t_{2222}=\Lambda_{2}, \\
t_{1122}=&\displaystyle\frac{1}{6}(\Lambda_{3}+\Lambda_{4}\rho^{2}+\Lambda_5\rho^2\cos2\theta),\\
t_{1112}=&\displaystyle\frac{1}{2}\Lambda_6\rho \cos\theta,\ t_{1222}=\displaystyle\frac{1}{2}\Lambda_7\rho \cos\theta.
\endaligned\end{equation}
That is, $\Lambda_1>0$, $\Lambda_2>0$,  the quartic form \begin{equation}\label{eq:8}
	\aligned
	\mathcal{T}(\rho,\theta){\bf x}^4=& \Lambda_1x_1^4+\Lambda_2x_2^4+(\Lambda_3+\Lambda_4\rho^2+\Lambda_5\rho^2\cos2\theta)x_1^2x_2^2\\
	&+2(\rho\Lambda_6\cos\theta) x_1^3x_2 +2(\rho\Lambda_7\cos\theta )x_1x_2^3 \geq0,
	\endaligned\end{equation}
for all $x_1\geq0, x_2\geq0$ if and only if
$$\aligned (\textbf{a})\ &\Lambda_6=\Lambda_7=0, \Lambda_3+2\sqrt{\Lambda_1\Lambda_2}\geq0, \Lambda_3+\Lambda_4-|\Lambda_5|+2\sqrt{\Lambda_1\Lambda_2}\geq0;\\
(\textbf{b})\	
& \Delta\geq0, \Lambda_3+2\sqrt{\Lambda_1\Lambda_2}\geq0, \Lambda_3+\Lambda_4-\Lambda_5+2\sqrt{\Lambda_1\Lambda_2}\geq0,\\ &|\Lambda_6\sqrt{\Lambda_2}-\Lambda_7\sqrt{\Lambda_1}|\leq2\sqrt{\Lambda_1\Lambda_2(\Lambda_3+\Lambda_4+\Lambda_5)+2\Lambda_1\Lambda_2\sqrt{\Lambda_1\Lambda_2}},\\
&\mbox{(i) }-2\sqrt{\Lambda_1\Lambda_2}\leq \Lambda_3+\Lambda_4+\Lambda_5 \leq6\sqrt{\Lambda_1\Lambda_2}, \\
&\mbox{(ii) }\Lambda_3+\Lambda_4+\Lambda_5>6\sqrt{\Lambda_1\Lambda_2}\mbox{ and }\\
&|\Lambda_6\sqrt{\Lambda_2}+\Lambda_7\sqrt{\Lambda_1}| \leq 2\sqrt{\Lambda_1\Lambda_2(\Lambda_3+\Lambda_4+\Lambda_5)-2\Lambda_1\Lambda_2\sqrt{\Lambda_1\Lambda_2}},
\endaligned$$
where $\Delta=4(12\Lambda_1\Lambda_2-12\Lambda_6\Lambda_7+(\Lambda_3+\Lambda_4+\Lambda_5)^{2})^{3}-(72\Lambda_1\Lambda_2(\Lambda_3+\Lambda_4+\Lambda_5)+36\Lambda_6\Lambda_7(\Lambda_3+\Lambda_4+\Lambda_5)-2(\Lambda_3+\Lambda_4+\Lambda_5)^{3}-108\Lambda_1\Lambda_7^{2}-108\Lambda_6^{2}\Lambda_2)^{2}$.

\section{Vacuum stability of the general  2HDM potential}
\subsection{Sufficient conditions}

In this section, we mainly give the vacuum stability conditions of the 2HDM potential \eqref{eq:V}  with explicit CP violation. We rewrite the quartic part of such a 2HDM potential as follow
 \begin{equation}\label{eq:9}\aligned V_4(\Phi_1,\Phi_2)=&\sum\limits_{i,j,k,l=1}^{2}t_{ijkl}(\Phi_i^*\Phi_j)(\Phi_k^*\Phi_l)\\
 	=&\lambda_1(\Phi_1^*\Phi_1)^2+\lambda_2(\Phi_2^*\Phi_2)^2\\&+\lambda_3(\Phi_1^*\Phi_1)(\Phi_2^*\Phi_2)+\lambda_4(\Phi_1^*\Phi_2)(\Phi_2^*\Phi_1)\\
	&+\frac{\lambda_5}2(\Phi_1^*\Phi_2)^2+\frac{\lambda_5^*}2(\Phi_2^*\Phi_1)^2\\&+(\Phi_1^*\Phi_1)(\lambda_6\Phi_1^*\Phi_2+\lambda_6^*\Phi_2^*\Phi_1)\\
	&+(\Phi_2^*\Phi_2)(\lambda_7\Phi_1^*\Phi_2+\lambda_7^*\Phi_2^*\Phi_1).
	\endaligned\end{equation}

Let $\phi_i$=$|\Phi_i|=\sqrt{\Phi_i^*\Phi_i}$, the modulus of $\Phi_i$ for $i=1,2.$ Then  $$\Phi_1^*\Phi_2=\phi_1\phi_2 \rho e^{i\theta}\mbox{ and }\Phi_2^*\Phi_1=\phi_1\phi_2 \rho e^{-i\theta},$$
here $i^2=-1$ and $\rho\in[0,1]$ is the orbit space parameter \cite{GC,K2016,GK2005}.  Let $$\lambda_5=|\lambda_5| e^{i\varphi_5}, \lambda_6=|\lambda_6| e^{i\varphi_6}, \lambda_7=|\lambda_7| e^{i\varphi_7},$$ where $\varphi_k$ is argument of the complex number $\lambda_k$ ($k=5,6,7$). Then $$\lambda_5^*=|\lambda_5| e^{-i\varphi_5}, \lambda_6^*=|\lambda_6| e^{-i\varphi_6}, \lambda_7^*=|\lambda_7| e^{-i\varphi_7}.$$ So, we have
$$\aligned V_4(\Phi_1,\Phi_2)=& \lambda_1\phi_1^4+\lambda_2\phi_2^4+\lambda_3\phi_1^2\phi_2^2+\lambda_4\rho^2\phi_1^2\phi_2^2\\
&+\frac{|\lambda_5|}2(e^{i(\varphi_5+2\theta)}+e^{-i(\varphi_5+2\theta)})\phi_1^2\phi_2^2\rho^2\\
&+|\lambda_6|( e^{i(\varphi_6+\theta)}+e^{-i(\varphi_6+\theta)})\phi_1^3\phi_2\rho\\
&+|\lambda_7|( e^{i(\varphi_7+\theta)}+e^{-i(\varphi_7+\theta)})\phi_1\phi_2^3\rho\\
=&\lambda_1\phi_1^4+\lambda_2\phi_2^4+\lambda_3\phi_1^2\phi_2^2+\lambda_4\rho^2\phi_1^2\phi_2^2\\
&+|\lambda_5|\phi_1^2\phi_2^2\rho^2\cos(\varphi_5+2\theta)\\
&+2|\lambda_6|\phi_1^3\phi_2\rho\cos( \varphi_6+\theta)\\
&+2|\lambda_7|\phi_1\phi_2^3\rho\cos( \varphi_7+\theta)\\
=&\lambda_1\phi_1^4+\lambda_2\phi_2^4+\lambda_3\phi_1^2\phi_2^2+\lambda_4\rho^2\phi_1^2\phi_2^2\\
&+|\lambda_5|\phi_1^2\phi_2^2\rho^2(\cos\varphi_5\cos2\theta-\sin\varphi_5\sin2\theta)\\
&+2|\lambda_6|\phi_1^3\phi_2\rho(\cos\varphi_6\cos\theta-\sin\varphi_6\sin\theta)\\
&+2|\lambda_7|\phi_1\phi_2^3\rho(\cos\varphi_7\cos\theta-\sin\varphi_7\sin\theta).
\endaligned$$
Obviously, $\textbf{Re}\lambda_k=|\lambda_k|\cos\varphi_k, \textbf{Im}\lambda_k=|\lambda_k|\sin\varphi_k, (k=5,6,7).$ Then, noticing $\sin2\theta=2\sin\theta\cos\theta$, we have
\begin{equation}\label{eq:10}
	\aligned
	V_4(\Phi_1,\Phi_2)=& \lambda_1\phi_1^4+\lambda_2\phi_2^4+(\lambda_3+\lambda_4\rho^2+\textbf{Re}\lambda_5\rho^2\cos2\theta)\phi_1^2\phi_2^2\\
	&-2(\rho\textbf{Im}\lambda_6\sin\theta)\phi_1^3\phi_2 -2(\rho\textbf{Im}\lambda_7\sin\theta)\phi_2^3\phi_1 \\
	&-2\rho\phi_1\phi_2(\textbf{Im}\lambda_5\rho\phi_1\phi_2\sin\theta\cos\theta\\
	&\ \ \ \ \ \ \ \ \ \ \ -\textbf{Re}\lambda_6\phi_1^2\cos\theta-\textbf{Re}\lambda_7\phi_2^2\cos\theta)\\
	=&V_4'(\phi_1,\phi_2)+V_4''(\phi_1,\phi_2),
	\endaligned \end{equation}
where \begin{equation}\label{eq:11}
	\aligned
	V_4'(\Phi_1,\Phi_2)=& \lambda_1\phi_1^4+\lambda_2\phi_2^4+(\lambda_3+\lambda_4\rho^2+\textbf{Re}\lambda_5\rho^2\cos2\theta)\phi_1^2\phi_2^2\\
	&-2(\rho\textbf{Im}\lambda_6\sin\theta)\phi_1^3\phi_2 -2(\rho\textbf{Im}\lambda_7\sin\theta)\phi_2^3\phi_1, \\
	V_4''(\Phi_1,\Phi_2)=&2(\rho\cos\theta)[\textbf{Re}\lambda_6\phi_1^2-(\rho\sin\theta)\textbf{Im}\lambda_5\phi_1\phi_2+\textbf{Re}\lambda_7\phi_2^2]\phi_1\phi_2.
	\endaligned\end{equation}

Applying the co-positivity of a real tensor \eqref{eq:7} with $$\Lambda_i=\lambda_i\  (i=1,2,3,4), \Lambda_k=\textbf{Re}\lambda_k\  (k=5,6,7)$$ to obtain that $\lambda_1>0, \lambda_2>0$,
\begin{equation}\label{eq:12}V_4'(\phi_1,\phi_2)\geq0\mbox{ for all } \phi_1,\phi_2\end{equation}
 if and only if
$$\aligned (\textbf{I})\ &\textbf{Im}\lambda_6=\textbf{Im}\lambda_7=0, \lambda_3+2\sqrt{\lambda_1\lambda_2}\geq0,\\ &\lambda_3+\lambda_4-|\textbf{Re}\lambda_5|+2\sqrt{\lambda_1\lambda_2}\geq0;\\
(\textbf{II})\	
& \textbf{Im}\lambda_6\ne0\mbox{ or }\textbf{Im}\lambda_7\ne0, \Delta\geq0, \\&\lambda_3+2\sqrt{\lambda_1\lambda_2}\geq0, \lambda_3+\lambda_4+\textbf{Re}\lambda_5+2\sqrt{\lambda_1\lambda_2}\geq0\\ &|\textbf{Im}\lambda_6\sqrt{\lambda_2}-\textbf{Im}\lambda_7\sqrt{\lambda_1}|\leq2\sqrt{\lambda_1\lambda_2(\lambda_3+\lambda_4-\textbf{Re}\lambda_5)+2\lambda_1\lambda_2\sqrt{\lambda_1\lambda_2}},\\
&\mbox{(i) }-2\sqrt{\lambda_1\lambda_2}\leq \lambda_3+\lambda_4-\textbf{Re}\lambda_5 \leq6\sqrt{\lambda_1\lambda_2}, \\
&\mbox{(ii) }\lambda_3+\lambda_4-\textbf{Re}\lambda_5>6\sqrt{\lambda_1\lambda_2}\mbox{ and }\\
&|\textbf{Im}\lambda_6\sqrt{\lambda_2}+\textbf{Im}\lambda_7\sqrt{\lambda_1}| \leq 2\sqrt{\lambda_1\lambda_2(\lambda_3+\lambda_4-\textbf{Re}\lambda_5)-2\lambda_1\lambda_2\sqrt{\lambda_1\lambda_2}},
\endaligned$$
where  $\Delta=4(12\lambda_1\lambda_2-12\textbf{Im}\lambda_6\cdot\textbf{Im}\lambda_7+(\lambda_3+\lambda_4-\textbf{Re}\lambda_5)^{2})^{3}-(72\lambda_1\lambda_2(\lambda_3+\lambda_4-\textbf{Re}\lambda_5)+36\textbf{Im}\lambda_6\cdot\textbf{Im}\lambda_7(\lambda_3+\lambda_4-\textbf{Re}\lambda_5)-2(\lambda_3+\lambda_4-\textbf{Re}\lambda_5)^{3}-108\lambda_1(\textbf{Im}\lambda_7)^{2}-108(\textbf{Im}\lambda_6)^{2}\lambda_2)^{2}$.\\

After making simple calculations ($\sin\theta\ne0$), we have
\begin{equation}\label{eq:13}V_4''(\phi_1,\phi_2)\geq0\mbox{ for all } \phi_1,\phi_2\end{equation}
if and only if  $$	\textbf{Re}\lambda_6\phi_1^2\cos\theta-	(\rho\cos\theta\sin\theta)\textbf{Im}\lambda_5\phi_1\phi_2 +\textbf{Re}\lambda_7\phi_2^2\cos\theta\geq0.$$

By the co-positivity of a real matrix \eqref{eq:6} with $$\mu_{11}=\textbf{Re}\lambda_6\cos\theta, \mu_{12}=-\dfrac12(\rho\cos\theta\sin\theta)\textbf{Im}\lambda_5, \mu_{22}=\textbf{Re}\lambda_7\cos\theta, $$ we obtain that $$\textbf{Re}\lambda_6\phi_1^2\cos\theta-	(\rho\cos\theta\sin\theta)\textbf{Im}\lambda_5\phi_1\phi_2 +\textbf{Re}\lambda_7\phi_2^2\cos\theta\geq0\mbox{ for all } \phi_1,\phi_2$$
if and only if 
$$ \textbf{Re}\lambda_6\cos\theta\geq0,\textbf{Re}\lambda_7\cos\theta\geq0, -(\rho\cos\theta\sin\theta)\textbf{Im}\lambda_5+2\sqrt{\textbf{Re}\lambda_6\textbf{Re}\lambda_7}|\cos\theta|\geq0.$$
Since  both $\rho\in[0,1]$ and $ \theta\in[0,2\pi]$ are arbitrary, then we have
$$\pm\textbf{Re}\lambda_6 \geq0,\pm \textbf{Re}\lambda_7\ge0, -(\textbf{Im}\lambda_5\sin\theta)\dfrac{\cos\theta}{|\cos\theta|}+2\sqrt{\textbf{Re}\lambda_6\textbf{Re}\lambda_7}\ge0,$$
and hence,  
$$\textbf{Re}\lambda_6=\textbf{Re}\lambda_7=0, \pm\textbf{Im}\lambda_5+2\sqrt{\textbf{Re}\lambda_6\textbf{Re}\lambda_7}\ge0.$$
So, we get the conclusion that $$V_4''(\Phi_1,\Phi_2)\geq0\mbox{ for all } \Phi_1,\Phi_2$$
if and only if
$$ \textbf{Re}\lambda_6=\textbf{Re}\lambda_7=\textbf{Im}\lambda_5=0.\leqno{(\textbf{III})}$$

In summary, we prove the analytic conditions (\textbf{I}), (\textbf{II}) and (\textbf{III}) assure the vacuum stability of the 2HDM potential with explicit CP violation. At the same time, we also obtain  the semi-positive definiteness of a 4th-order 2-dimensional complex tensor $\mathcal{T}=(t_{ijkl})$  definited by the Eq. \eqref{eq:T}.\\

In term of  Eq. \eqref{eq:07}, we also may obtain a stronger sufficient condition. That is,
$V_4(\Phi_1,\Phi_2)\geq0\mbox{ for all } \Phi_1,\Phi_2$
if  for all $\rho\in[0,1]$ and all $\theta\in[0,2\pi]$, $$\aligned	& \beta(\theta)=2\rho|\lambda_6|\cos(\varphi_6+\theta)+4\sqrt[4]{\lambda_1^3\lambda_2}\geq0,\\ &\gamma(\theta)=2\rho|\lambda_7|\cos(\varphi_7+\theta)+4\sqrt[4]{\lambda_1\lambda_2^3}\geq0,\\
& (\lambda_3+\lambda_4\rho^2+|\lambda_5|\rho^2\cos(\varphi_5+2\theta))-6\sqrt{\lambda_1\lambda_2}+2\sqrt{\beta(\theta)\cdot\gamma(\theta)}\geq0.
\endaligned$$
It is obvious that the above inequality system  comes true if
$$\aligned	& \beta=2\sqrt[4]{\lambda_1^3\lambda_2}-|\lambda_6|\geq0, \gamma=2\sqrt[4]{\lambda_1\lambda_2^3}-|\lambda_7|\geq0,\\
& \lambda_3-6\sqrt{\lambda_1\lambda_2}+4\sqrt{\beta\gamma}\geq0,\\
& \lambda_3+\lambda_4-|\lambda_5|-6\sqrt{\lambda_1\lambda_2}+4\sqrt{\beta\gamma}\geq0.
\endaligned \leqno{(\textbf{IV})}$$

Similarly, $V_4'(\phi_1,\phi_2)\geq0\mbox{ for all } \phi_1,\phi_2$
if
$$\aligned	& \beta'=2\sqrt[4]{\lambda_1^3\lambda_2}-|\textbf{Im}\lambda_6|\geq0, \gamma'=2\sqrt[4]{\lambda_1\lambda_2^3}-|\textbf{Im}\lambda_7|\geq0,\\
& \lambda_3-6\sqrt{\lambda_1\lambda_2}+4\sqrt{\beta'\gamma'}\geq0,\\
& \lambda_3+\lambda_4-|\textbf{Re}\lambda_5|-6\sqrt{\lambda_1\lambda_2}+4\sqrt{\beta'\gamma'}\geq0.
\endaligned \leqno{(\textbf{IV}')}$$

\begin{remark} \em Four analytical sufficient conditions are  following:
	\begin{itemize}
		\item[(1)]  $\textbf{Re}\lambda_6=\textbf{Re}\lambda_7=0$,  the conditions (\textbf{I}) and  (\textbf{III});
		\item[(2)]  $\textbf{Re}\lambda_6\ne0$ or $\textbf{Re}\lambda_7\ne0$,  the conditions (\textbf{II}) and  (\textbf{III});
		\item[(3)]   the conditions $(\textbf{IV}')$ and  (\textbf{III});
		\item[(4)]   the conditions (\textbf{IV}).
	\end{itemize}
	
	These analytical sufficient conditions are obtained from the real and imaginary parts of complex numbers, not only dependent of the norm.  So they are different from the ones of Bahl et.al.\cite{BCCIW}. For example,  $$\aligned	 \lambda_1=&\lambda_2=1, \lambda_3=6, \lambda_4=2,  \lambda_5=-1-i, \lambda_6= \lambda_7=-1-\sqrt3i.\\
	\mbox{Obviously, } &|\lambda_6|=|\lambda_7|=2, \beta=\gamma=0,  \lambda_3-6\sqrt{\lambda_1\lambda_2}+4\sqrt{\beta\gamma}=0,\\ &\lambda_3+\lambda_4-|\lambda_5|-6\sqrt{\lambda_1\lambda_2}+4\sqrt{\beta\gamma}=2-\sqrt2>0.
	\endaligned $$
	That is, these parameters meet the condition (\textbf{IV}), which means $V_4(\Phi_1,\Phi_2)\geq0$.  However, they can't satisfy the condition Eq. (5.20) of Bahl et.al.\cite{BCCIW}, i.e.
	$$\aligned	& 3\sqrt{\lambda_1\lambda_2}-(\lambda_3+|\lambda_4|+|\lambda_5|)=3-(6+2+\sqrt{2})<0,\\ &\sqrt{\lambda_1\lambda_2}+\lambda_3-(|\lambda_4|+|\lambda_5|+4\left|\lambda_6\sqrt[4]{\dfrac{\lambda_2}{\lambda_1}}\right|)=1+6-(2+\sqrt{2}+8)<0,\\
	&\sqrt{\lambda_1\lambda_2}+\lambda_3-(|\lambda_4|+|\lambda_5|+4\left|\lambda_7\sqrt[4]{\dfrac{\lambda_1}{\lambda_2}}\right|)=1+6-(2+\sqrt{2}+8)<0.
	\endaligned $$
	Also see the example in the introduce.  
\end{remark}

\subsection{Sufficient and necessary conditions}
In this subsection, $V_4(\Phi_1,\Phi_2)$ is rewritten as follows ($\lambda_1>0,\lambda_2>0$),  \begin{equation}\label{eq:14}
	\aligned
	V_4&(\Phi_1,\Phi_2)= A\rho^2+B\rho+C=f(\rho),\\
	A&=a\phi_1^2\phi_2^2,\ B=b\phi_1\phi_2, \ C=\lambda_1\phi_1^4+\lambda_2\phi_2^4+\lambda_3\phi_1^2\phi_2^2,\\
	a&=\lambda_4-\textbf{Re}\lambda_5+2(\textbf{Re}\lambda_5\cos\theta-\textbf{Im}\lambda_5\sin\theta)\cos\theta\\
	&=\lambda_4+|\lambda_5|\cos(\varphi_5+2\theta),\\
	b&=2(\textbf{Re}\lambda_6\phi_1^2+\textbf{Re}\lambda_7\phi_2^2)\cos\theta-2(\textbf{Im}\lambda_6\phi_1^2+\textbf{Im}\lambda_7\phi_2^2)\sin\theta\\
	&=2|\lambda_6|\phi_1^2\cos(\varphi_6+\theta)+2|\lambda_7|\phi_2^2\cos(\varphi_7+\theta).
	\endaligned\end{equation}

The quadratic function $f(\rho)$ is non-negative about a variable $\rho\in[0,1]$ if and only if its minimum is non-negative in the interval $[0,1]$, and so, its function value is non-negative at the  boundary points $\rho=0,1$ and the unique  minimum point $\rho_0=-\frac{B}{2A}\in[0,1](A>0)$. That is,
$$f(\rho)\geq0 \Leftrightarrow \begin{cases}
	f(-\frac{B}{2A})=\dfrac{4AC-B^2}{4A}\geq0, -\frac{B}{2A}\in[0,1],\\
	f(0)\geq0, \\
	f(1)\geq0.
\end{cases} $$
The graph-like of $f(\rho)$ is  as shown below:
\begin{center}\begin{tikzpicture}
		\begin{axis}[
			xlabel={$\rho$}, ylabel={$y$},
			xmin=-2.1, xmax=6.3,
			ymin=-0.3, ymax=2.1,
			xtick={-2, -1,  0, 0.5, 1, 2},
			xticklabels={$-2$, $-1$, $0$, $\frac12$, $1$, $2$},
			line width=1pt,
			axis lines=center,
			]
		\addplot[smooth,domain=-2:3, red!70]{(x-0.5)^2};
		\addlegendentry{\tiny $0<-\frac{B}{2A}<1, A>0 $}
		\addplot[smooth,domain=-4:2, blue!70]{(x+0.52)^2+0.1};
		\addlegendentry{\tiny $-\frac{B}{2A}<0, A>0 $}
		\addplot[smooth,domain=-2:4]{(x-1.6)^2-0.2};
		\addlegendentry{\tiny  $-\frac{B}{2A}>1, A>0 $}
		\addplot[smooth,domain=-4:4, green!70]{-(x-1.3)^2+2};
		\addlegendentry{\tiny $-\frac{B}{2A}>\frac12, A<0 $}
		\addplot[smooth,domain=-3:4, teal!70]{-(x+0.2)^2+2};
		\addlegendentry{\tiny  $-\frac{B}{2A}<\frac12, A<0 $}
		\addplot[dashed, domain = 0:4, pink!70] (0.5,{x});
		\addlegendentry{\tiny  $x=\frac12 $}	
			\addplot[dashed, domain = 0:4, cyan!70] (0,{x});
			\addplot[dashed, domain = 0:4, cyan!70] (1,{x});
		\end{axis}
	\end{tikzpicture}
	
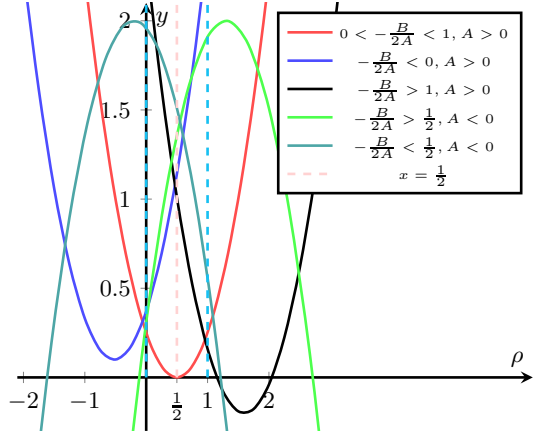
Figure 1: Graph-like of $f(\rho)$
\end{center}

\begin{proposition}\label{pro:1}\em
	$V_4(\Phi_1,\Phi_2)\geq0$  if and only if
	$$\begin{cases}
		4AC-B^2\geq0, &-2A\leq B\leq 0;\\
			C\geq0; &\\
			A+B+C\geq0. &
	\end{cases}$$
\end{proposition}
It is obvious that if $\lambda_6=\lambda_7=0$, then $B=0,$ the symmetry axis $-\frac{B}{2A}=0$, and hence,
$V_4(\Phi_1,\Phi_2)\geq0$  if and only if
$$
C\geq0\mbox{ and } A+C\geq0.
$$
For the 2HDM with $\mathbb{Z}_2$ symmetry \cite{N2020},  the quartic part of the general 2HDM  scalar potential is  \begin{equation}\label{eq:15}\aligned V_4^{\mathbb{Z}_2}(\Phi_1,\Phi_2)
	=&\lambda_1(\Phi_1^*\Phi_1)^2+\lambda_2(\Phi_2^*\Phi_2)^2\\&+\lambda_3(\Phi_1^*\Phi_1)(\Phi_2^*\Phi_2)+\lambda_4(\Phi_1^*\Phi_2)(\Phi_2^*\Phi_1)\\
	&+\frac{\lambda_5}2(\Phi_1^*\Phi_2)^2+\frac{\lambda_5^*}2(\Phi_2^*\Phi_1)^2.
	\endaligned\end{equation}
Therefore,  $V_4^{\mathbb{Z}_2}(\Phi_1,\Phi_2)\geq0$  if and only if
$$\begin{cases}
C=\lambda_1\phi_1^4+\lambda_2\phi_2^4+\lambda_3\phi_1^2\phi_2^2\geq0\\
A+C=\lambda_1\phi_1^4+\lambda_2\phi_2^4+(\lambda_3+\lambda_4+|\lambda_5|\cos(\varphi_5+2\theta))\phi_1^2\phi_2^2\geq0.
\end{cases}
$$
It is clear that $C\geq0\Leftrightarrow  \lambda_3+2\sqrt{\lambda_1\lambda_2}\geq0,$
$$\aligned A+C\geq0\Leftrightarrow&  \lambda_3+\lambda_4+|\lambda_5|\cos(\varphi_5+2\theta)+2\sqrt{\lambda_1\lambda_2}\geq0, \forall \theta\in[0,2\pi]\\
\Leftrightarrow&  \lambda_3+\lambda_4-|\lambda_5|+2\sqrt{\lambda_1\lambda_2}\geq0.\endaligned$$
\begin{corollary}\label{cor:2} \em
	$V_4^{\mathbb{Z}_2}(\Phi_1,\Phi_2)\geq0$  if and only if $\lambda_1\geq0,\lambda_2\geq0$,
$$ \lambda_3+2\sqrt{\lambda_1\lambda_2}\geq0, \lambda_3+\lambda_4-|\lambda_5|+2\sqrt{\lambda_1\lambda_2}\geq0.	\leqno{(\textbf{V})}$$
\end{corollary}
This condition (\textbf{V}) is well-known for the inert doublet model \cite{K2012,DM,GK2005,K,MMNN}.
\subsection{Necessary conditions}
In this subsection, $V_4(\Phi_1,\Phi_2)$ is rewritten as follows ($\lambda_1>0,\lambda_2>0$),
\begin{equation}\label{eq:17}
	\aligned
	V_4(\Phi_1,&\Phi_2)=f(\rho)= A\rho^2+B\rho+C,\\
	f(1)=& A+B+C\\
	=&\lambda_1\phi_1^4+\lambda_2\phi_2^4+(\lambda_3+\lambda_4-\textbf{Re}\lambda_5+2\textbf{Re}\lambda_5\cos^2\theta)\phi_1^2\phi_2^2\\
	&+2(\textbf{Re}\lambda_6\cos\theta)\phi_1^3\phi_2 +2(\textbf{Re}\lambda_7\cos\theta)\phi_2^3\phi_1 \\
	&-2(\sin\theta)[(\cos\theta)\textbf{Im}\lambda_5\phi_1\phi_2\\
	&\ \ \ \ \ \ \ \ \ \ \ \ \ \ \ \  +\textbf{Im}\lambda_6\phi_1^2+\textbf{Im}\lambda_7\phi_2^2]\phi_1\phi_2.
	\endaligned\end{equation}
Obviously, we have
$$f(\rho)\geq0, \mbox{ for all }\rho\in[0,1] \Rightarrow
	f(0)\geq0,
	f(1)\geq0.
 $$
Then $V_4(\Phi_1,\Phi_2)\geq0$ implies that $f(0)=C\geq0$, which is equvalent to $$\lambda_3+2\sqrt{\lambda_1\lambda_2}\geq0.\leqno{(\textbf{VI})}$$
This is a necessary condition of the vacuum stability of the general 2HDM potential. Clearly, the other necessary condition is some conditions such that $f(1)=A+B+C\geq0$. By Eq. \eqref{eq:17}, it is known that $A+B+C$ may be regarded as a quartic form with two parameters $t=\sin \theta$ and $s=\cos\theta$ with $s^2+t^2=1.$ So, when $s=\sin \theta=0$ and $t=\cos\theta=\pm1$,   the inequality  $$\lambda_1\phi_1^4+\lambda_2\phi_2^4+(\lambda_3+\lambda_4+\textbf{Re}\lambda_5)\phi_1^2\phi_2^2
\pm2\textbf{Re}\lambda_6\phi_1^3\phi_2 \pm2\textbf{Re}\lambda_7\phi_2^3\phi_1\geq0$$
 if and only if (using Eq.\eqref{eq:06})
\begin{flushleft} (1) $\Delta\leq0$, $\textbf{Re}\lambda_6\sqrt{\lambda_2}+\textbf{Re}\lambda_7\sqrt{\lambda_1}>0$;  \end{flushleft} \begin{flushleft} (2) $\textbf{Re}\lambda_6\geq0$, $\textbf{Re}\lambda_7\geq0$, $\lambda_3+\lambda_4+\textbf{Re}\lambda_5+2\sqrt{\lambda_1\lambda_2}\geq0$; \end{flushleft}
\begin{flushleft}  (3) $\Delta\geq0$,
	$|\textbf{Re}\lambda_6\sqrt{\lambda_2}-\textbf{Re}\lambda_7\sqrt{\lambda_1}|\leq2\sqrt{\lambda_1\lambda_2(\lambda_3+\lambda_4+\textbf{Re}\lambda_5)+2\lambda_1\lambda_2\sqrt{\lambda_1\lambda_2}}$,
\end{flushleft}
\begin{flushleft}
	\qquad(i) $-2\sqrt{\lambda_1\lambda_2}\leq \lambda_3+\lambda_4+\textbf{Re}\lambda_5 \leq6\sqrt{\lambda_1\lambda_2}$,
\end{flushleft}
\begin{flushleft}
	\qquad(ii) $\lambda_3+\lambda_4+\textbf{Re}\lambda_5>6\sqrt{\lambda_1\lambda_2}$, \end{flushleft}
\begin{flushleft}
$\textbf{Re}\lambda_6\sqrt{\lambda_2}+\textbf{Re}\lambda_7\sqrt{\lambda_1} \geq -2\sqrt{\lambda_1\lambda_2(\lambda_3+\lambda_4+\textbf{Re}\lambda_5)-2\lambda_1\lambda_2\sqrt{\lambda_1\lambda_2}}$.
\end{flushleft}
and \begin{flushleft} $(1')$ $\Delta\leq0$, $-\textbf{Re}\lambda_6\sqrt{\lambda_2}-\textbf{Re}\lambda_7\sqrt{\lambda_1}>0$;  \end{flushleft} \begin{flushleft} $(2')$ $-\textbf{Re}\lambda_6\geq0$, $-\textbf{Re}\lambda_7\geq0$, $\lambda_3+\lambda_4+\textbf{Re}\lambda_5+2\sqrt{\lambda_1\lambda_2}\geq0$; \end{flushleft}
\begin{flushleft}  $(3')$ $\Delta\geq0$,
	$|-\textbf{Re}\lambda_6\sqrt{\lambda_2}+\textbf{Re}\lambda_7\sqrt{\lambda_1}|\leq2\sqrt{\lambda_1\lambda_2(\lambda_3+\lambda_4+\textbf{Re}\lambda_5)+2\lambda_1\lambda_2\sqrt{\lambda_1\lambda_2}}$,
\end{flushleft}
\begin{flushleft}
	\qquad$(i')$ $-2\sqrt{\lambda_1\lambda_2}\leq \lambda_3+\lambda_4+\textbf{Re}\lambda_5 \leq6\sqrt{\lambda_1\lambda_2}$,
\end{flushleft}
\begin{flushleft}
	\qquad$(ii')$ $\lambda_3+\lambda_4+\textbf{Re}\lambda_5>6\sqrt{\lambda_1\lambda_2}$, \end{flushleft}
\begin{flushleft}
	$-\textbf{Re}\lambda_6\sqrt{\lambda_2}-\textbf{Re}\lambda_7\sqrt{\lambda_1} \geq -2\sqrt{\lambda_1\lambda_2(\lambda_3+\lambda_4+\textbf{Re}\lambda_5)-2\lambda_1\lambda_2\sqrt{\lambda_1\lambda_2}}$.
\end{flushleft}
Which is equivalent to
 $$\aligned  &\textbf{Re}\lambda_6=\textbf{Re}\lambda_7=0, \lambda_3+\lambda_4+\textbf{Re}\lambda_5+2\sqrt{\lambda_1\lambda_2}\geq0;\\
& \textbf{Re}\lambda_6\ne0\mbox{ or }\textbf{Re}\lambda_7\ne0,  \Delta\geq0,\\
&|\textbf{Re}\lambda_6\sqrt{\lambda_2}-\textbf{Re}\lambda_7\sqrt{\lambda_1}|\leq2\sqrt{\lambda_1\lambda_2(\lambda_3+\lambda_4+\textbf{Re}\lambda_5)+2\lambda_1\lambda_2\sqrt{\lambda_1\lambda_2}}\\
& \quad(a)  -2\sqrt{\lambda_1\lambda_2}\leq \lambda_3+\lambda_4+\textbf{Re}\lambda_5 \leq6\sqrt{\lambda_1\lambda_2},\\
&\quad(b)  \lambda_3+\lambda_4+\textbf{Re}\lambda_5>6\sqrt{\lambda_1\lambda_2}, \\
& |\textbf{Re}\lambda_6\sqrt{\lambda_2}+\textbf{Re}\lambda_7\sqrt{\lambda_1}| \leq 2\sqrt{\lambda_1\lambda_2(\lambda_3+\lambda_4+\textbf{Re}\lambda_5)-2\lambda_1\lambda_2\sqrt{\lambda_1\lambda_2}}.\endaligned$$
So, a necessary condition of $V_4(\Phi_1,\Phi_2)\geq0$ is $$\lambda_3+\lambda_4+\textbf{Re}\lambda_5+2\sqrt{\lambda_1\lambda_2}\geq0.\leqno{(\textbf{VII})}$$
Similarly, if $t=\cos \theta=0$ and $s=\sin\theta=\pm1$,   the inequality  $$\lambda_1\phi_1^4+\lambda_2\phi_2^4+(\lambda_3+\lambda_4-\textbf{Re}\lambda_5)\phi_1^2\phi_2^2
\pm2\textbf{Im}\lambda_6\phi_1^3\phi_2 \pm2\textbf{Im}\lambda_7\phi_2^3\phi_1\geq0$$
  is equivalent to
$$\aligned &\textbf{Im}\lambda_6=\textbf{Im}\lambda_7=0, \lambda_3+\lambda_4-\textbf{Re}\lambda_5+2\sqrt{\lambda_1\lambda_2}\geq0;\\
&\textbf{Im}\lambda_6\ne0\mbox{ or }\textbf{Im}\lambda_7\ne0, \Delta'\geq0,\\
&|\textbf{Im}\lambda_6\sqrt{\lambda_2}-\textbf{Im}\lambda_7\sqrt{\lambda_1}|\leq2\sqrt{\lambda_1\lambda_2(\lambda_3+\lambda_4-\textbf{Re}\lambda_5)+2\lambda_1\lambda_2\sqrt{\lambda_1\lambda_2}}\\
& \quad(a')  -2\sqrt{\lambda_1\lambda_2}\leq \lambda_3+\lambda_4-\textbf{Re}\lambda_5 \leq6\sqrt{\lambda_1\lambda_2},\\
&\quad(b')  \lambda_3+\lambda_4-\textbf{Re}\lambda_5>6\sqrt{\lambda_1\lambda_2}, \\
& |\textbf{Im}\lambda_6\sqrt{\lambda_2}+\textbf{Im}\lambda_7\sqrt{\lambda_1}| \leq 2\sqrt{\lambda_1\lambda_2(\lambda_3+\lambda_4-\textbf{Re}\lambda_5)-2\lambda_1\lambda_2\sqrt{\lambda_1\lambda_2}},\endaligned$$
where $\Delta'=4(12\lambda_1\lambda_2-12\textbf{Im}\lambda_6\cdot\textbf{Im}\lambda_7+(\lambda_3+\lambda_4-\textbf{Re}\lambda_5)^{2})^{3}-(72\lambda_1\lambda_2(\lambda_3+\lambda_4-\textbf{Re}\lambda_5)+36\textbf{Im}\lambda_6\cdot\textbf{Im}\lambda_7(\lambda_3+\lambda_4-\textbf{Re}\lambda_5)-2(\lambda_3+\lambda_4-\textbf{Re}\lambda_5)^{3}-108\lambda_1(\textbf{Im}\lambda_7)^{2}-108(\textbf{Im}\lambda_6)^{2}\lambda_2)^{2}$. 
So, a necessary condition of $V_4(\Phi_1,\Phi_2)\geq0$ is $$\lambda_3+\lambda_4-\textbf{Re}\lambda_5+2\sqrt{\lambda_1\lambda_2}\geq0.\leqno{(\textbf{VIII})}$$
Applying the Corollary 3.1 of Song and Qi \cite{SQ2021} to $V_4(\Phi_1,\Phi_2)$,
$$\aligned V_4(\Phi_1,\Phi_2)
=&\lambda_1\phi_1^4+\lambda_2\phi_2^4+[\lambda_3+\lambda_4\rho^2 +|\lambda_5|\rho^2\cos(\varphi_5+2\theta)]\phi_1^2\phi_2^2\\
&+2|\lambda_6|\phi_1^3\phi_2\rho\cos( \varphi_6+\theta)+2|\lambda_7|\phi_1\phi_2^3\rho\cos( \varphi_7+\theta),
\endaligned$$
we obtain that $V_4(\Phi_1,\Phi_2)>0$ implies that for all $\rho\in[0,1]$ and all $\theta\in[0,2\pi]$,
$$\aligned 0< &2\left( \frac14\times 2|\lambda_6|\rho\cos( \varphi_6+\theta)\right)\sqrt{\lambda_2}+ 2\left(\frac14\times 2|\lambda_7|\rho\cos( \varphi_7+\theta)\right)\sqrt{\lambda_1}\\
&\ \ +\left(3\times\frac16(\lambda_3+\lambda_4\rho^2+|\lambda_5|\rho^2\cos(\varphi_5+2\theta))+\sqrt{\lambda_1\lambda_2}\right)\sqrt[4]{\lambda_1\lambda_2}\\
=&|\lambda_6|\sqrt{\lambda_2}\rho\cos( \varphi_6+\theta) +|\lambda_7|\sqrt{\lambda_1}\rho\cos( \varphi_7+\theta)\\
&+\frac12\left(\lambda_3+\lambda_4\rho^2+|\lambda_5|\rho^2\cos(\varphi_5+2\theta)+2\sqrt{\lambda_1\lambda_2}\right)\sqrt[4]{\lambda_1\lambda_2}\\
=&\textbf{Re}\lambda_6\sqrt{\lambda_2}\rho\cos\theta +\textbf{Re}\lambda_7\sqrt{\lambda_2}\rho\cos\theta\\
&+\frac12\left(\lambda_3+\lambda_4\rho^2+\textbf{Re}\lambda_5\rho^2\cos2\theta+2\sqrt{\lambda_1\lambda_2}\right)\sqrt[4]{\lambda_1\lambda_2}\\
&-\textbf{Im}\lambda_6\sqrt{\lambda_2}\rho\sin\theta -\textbf{Im}\lambda_7\sqrt{\lambda_1}\rho\sin\theta-\frac12\textbf{Im}\lambda_5\sqrt[4]{\lambda_1\lambda_2}\rho^2\sin2\theta, \endaligned$$
and then, $\rho=1$ and  $\theta=0\mbox{ or }\pi \mbox{ or }\dfrac\pi2\mbox{ or }\dfrac{3\pi}2$, the above inequality must holds also. That is,
$$\aligned (\lambda_3+\lambda_4+&\textbf{Re}\lambda_5+\left.2\sqrt{\lambda_1\lambda_2}\right)\sqrt[4]{\lambda_1\lambda_2}\pm2\left(\textbf{Re}\lambda_6\sqrt{\lambda_2} +\textbf{Re}\lambda_7\sqrt{\lambda_1}\right)>0\\
(\lambda_3+\lambda_4-&\textbf{Re}\lambda_5+\left.2\sqrt{\lambda_1\lambda_2}\right)\sqrt[4]{\lambda_1\lambda_2}\mp2\left(\textbf{Im}\lambda_6\sqrt{\lambda_2} +\textbf{Im}\lambda_7\sqrt{\lambda_1}\right)>0.\endaligned $$
or equivalently, $$\aligned (\lambda_3+\lambda_4+&\textbf{Re}\lambda_5+\left.2\sqrt{\lambda_1\lambda_2}\right)\sqrt[4]{\lambda_1\lambda_2}\\ &\ \ -2\left|\textbf{Re}\lambda_6\sqrt{\lambda_2} +\textbf{Re}\lambda_7\sqrt{\lambda_1}\right|>0\\
(\lambda_3+\lambda_4-&\textbf{Re}\lambda_5+\left.2\sqrt{\lambda_1\lambda_2}\right)\sqrt[4]{\lambda_1\lambda_2}\\
&\ \ -2\left|\textbf{Im}\lambda_6\sqrt{\lambda_2} +\textbf{Im}\lambda_7\sqrt{\lambda_1}\right|>0.\endaligned \leqno{(\textbf{IX})}$$
Clearly,  the condition (\textbf{VI}) (strict inequality) is obtained if $\rho=0$.\\

In summary, the conditions  (\textbf{VI}),  (\textbf{VII}), (\textbf{VIII}) and  (\textbf{IX}) are the necessary conditions of the vacuum stability of the general 2HDM potential.

\begin{remark}\em
	The conditions (\textbf{I}) and (\textbf{II}) obviuosly  meet (\textbf{VI})-(\textbf{VIII}) (necessary).  For (\textbf{IV}) and  (\textbf{IV}$'$), 
	$$\aligned	 \beta=2\sqrt[4]{\lambda_1^3\lambda_2}&-|\lambda_6|\geq0, \gamma=2\sqrt[4]{\lambda_1\lambda_2^3}-|\lambda_7|\geq0,\\
	 \lambda_3+2\sqrt{\lambda_1\lambda_2}&= \lambda_3-6\sqrt{\lambda_1\lambda_2}+4\sqrt{(2\sqrt[4]{\lambda_1^3\lambda_2})(2\sqrt[4]{\lambda_1\lambda_2^3})}\\&\geq\lambda_3-6\sqrt{\lambda_1\lambda_2}+4\sqrt{\beta\gamma}\geq0,\endaligned $$
	 i.e., $ \lambda_3+2\sqrt{\lambda_1\lambda_2}\geq0.$  So, the condition (VI) holds. Similarly, 
$$\aligned	& \lambda_3+\lambda_4-|\lambda_5|-6\sqrt{\lambda_1\lambda_2}+4\sqrt{(2\sqrt[4]{\lambda_1^3\lambda_2})(2\sqrt[4]{\lambda_1\lambda_2^3})}\\
	& \geq \lambda_3+\lambda_4-|\lambda_5|-6\sqrt{\lambda_1\lambda_2}+4\sqrt{\beta\gamma}\geq0,\endaligned $$
	i.e., $\lambda_3+\lambda_4-|\lambda_5|+2\sqrt{\lambda_1\lambda_2}\geq0.$  It follows from the inequalities $- |\lambda_5|\leq\textbf{Re}\lambda_5\leq |\lambda_5|$ that
	$$\aligned
	& \lambda_3+\lambda_4+\textbf{Re}\lambda_5+2\sqrt{\lambda_1\lambda_2}\geq\lambda_3+\lambda_4-|\lambda_5|+2\sqrt{\lambda_1\lambda_2}\geq0,\\
	&\lambda_3+\lambda_4-\textbf{Re}\lambda_5+2\sqrt{\lambda_1\lambda_2}\geq\lambda_3+\lambda_4-|\lambda_5|+2\sqrt{\lambda_1\lambda_2}\geq0. 
	\endaligned $$
So, the conditions (\textbf{VII}) and (\textbf{VIII}) hold.	Similarly, the condition (\textbf{IV}$'$) meets (\textbf{VI})-(\textbf{VIII}) also.  The condition (\textbf{IX}) is necessary condition of ``strict inequality", $V_4(\Phi_1,\Phi_2)>0$. 
\end{remark}

\section{Conclusions}
By means of the co-positive conditions of a 4th-order symmetry tensor,  several analytical sufficient conditions and necessary conditions are established for the vacuum stability of the general 2HDM potential,  respectively.  That is, \\
Four sufficient conditions: $ \begin{cases}
	 (1)\   (\textbf{I}) \mbox{ and } (\textbf{III});\\
	 (2)\  (\textbf{II})\mbox{ and } (\textbf{III});\\
	 (3)  \ (\textbf{IV}') \mbox{ and }  (\textbf{III});\\
	 (4) \  (\textbf{IV}).
\end{cases}$\\
Four necessary conditions:   (\textbf{VI}),  (\textbf{VII}), (\textbf{VIII}) and  (\textbf{IX}) .\\

A sufficient  and necessary condition is qualitatively showed for the vacuum stability of the general 2HDM potential, and then, applying it to derive the analytical necessary conditions for the vacuum stability of the general 2HDM potential. The vacuum stability condition (\textbf{V}) of the $\mathbb{Z}_2$ symmetry 2HDM potential is a special case.  
\begin{center}
		\begin{tikzpicture}[->, auto,node distance=7em, thick]
		
		\node[draw, rounded corners] (X) {\bf\color{magenta}\ I \ };
		\node [draw, rounded corners](V) [right of=X] {\bf\color{red}\ $V_4\geq0$ \ };
		\node[draw, rounded corners] (T) [above right of=V] {\bf \color{violet}\ V  \ };
		\node [draw, rounded corners](II) [above left of=V] {\bf\color{cyan}\ II \ };
		\node [draw, rounded corners](Y) [above  of=V] {\bf\color{teal}IV};
		\node[draw, rounded corners] (S) [below  of=V] {\bf \color{blue}VIII};
		\node [draw, rounded corners](Z) [right of=V] {\bf \color{blue}VI};
		\node[draw, rounded corners] (VI) [below right of=V] {\bf \color{blue}VII};
		\node[draw, rounded corners] (VIII) [below left of=V] {\bf \color{green}\ IX \ };
		\node[draw, rounded corners, above=of X, xshift=1.5cm, yshift=-0.5cm](I)  {\bf\color{purple}IV$'$};
		
		\draw (X) -- (V) node[midway, above] {\bf\color{magenta}III};
		\draw (I) -- (V) node[midway, above] {\bf\color{purple}III};
		\draw (V) -- (Z) node[midway, above] { };
		\draw (II) -- (V) node[midway, above] {\bf \color{cyan}III \ \ \ };
		\draw (Y) -- (V) node[midway, below] { };
		\draw (T.220) -- (V.60) node[midway, above] { };
		\draw (V.35) -- (T.245) node[midway, above] {\tiny \bf \color{violet}\ \ \ $\mathbb{Z}_2$ symmetry };
		\draw (V) -- (VIII) node[midway, above] {\tiny \bf \color{green} strict\ \ \ \ \  };
		\draw (V) -- (VIII) node[midway, below] {\tiny \bf \color{green} inequality};
    		\draw (V) -- (VI) node[midway, above] { };
		\draw (V) -- (S) node[midway, above] { };
	\end{tikzpicture}

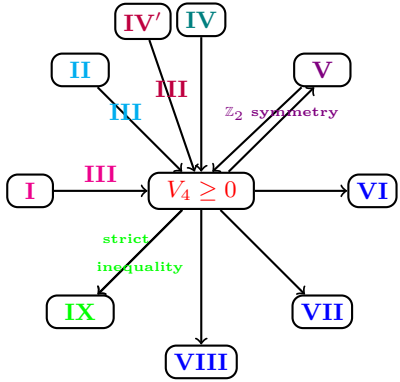
Figure 2: Analytical conditions and the vacuum stability \\ of the general 2HDM potential (``$\rightarrow$" stand for ``imply")
\end{center}
 
\section*{Competing interest}
The authors declare that they have no known competing financial interests or personal rela-tionships that could have appeared to influence the work reported in this paper.
\section*{Authors' contributions}
Yisheng Song is the sole author.
\section*{Availability of data and materials}
This manuscript has no associated data or the data will not be deposited. [Authors’ comment: This is a theoretical study and there are no external data associated with the manuscript].
\section*{Funding}
This  work was supported by
the National Natural Science Foundation of P.R. China (Grant No.12171064), by The team project of innovation leading talent in chongqing (No.CQYC20210309536) and by the Foundation of Chongqing Normal university (20XLB009).
\section*{Acknowledgements} The authors would like to express their sincere thanks to the  Editors and the anonymous referees for their constructive comments and valuable suggestions.


\begin{thebibliography}{}
\bibitem{L1973} T. Lee, A Theory of Spontaneous $T$ Violation, Phys. Rev. D 8,  1226(1973).
\bibitem{L1974} T. Lee, CP nonconservation and spontaneous symmetry breaking, Phys. Rep. 9(2) 143-177(1974).
\bibitem{SW} S. Weinberg, Gauge Theory of CP Nonconservation, Phys. Rev. Lett. 37, 657 (1976).
\bibitem{DM}N. G. Deshpande, E. Ma, Pattern of symmetry breaking with two Higgs doublets, Phys. Rev. D  18, 2574-2576(1978).
\bibitem{WW}Y.L. Wu and L. Wolfenstein, Sources of  CP Violation in the Two-Higgs-Doublet Model, Phys. Rev. Lett.  73, 1762(1994).
\bibitem{PW}A. Pilaftsis, C.E.M. Wagner, Higgs bosons in the minimal supersymmetric Standard Model with explicit CP violation, Nuclear Physics B 553, 3-42 (1999).
\bibitem{BFLRS} G.C. Branco, P.M. Ferreira, L. Lavoura, M.N. Rebelo, M. Sher, and J.P. Silva, Theory and phenomenology of two-Higgs-doublet models, Phys. Rep. 516, 1-102 (2012).
\bibitem{IS2015} I. P. Ivanov, J.P. Silva, Tree-level metastability bounds for the most general two Higgs doublet model. Phys. Rev. D  92, 055017(2015).
\bibitem{BS}F.J. Botella, J.P. Silva, Jarlskog-like invariants for theories with scalars and fermions, Phys. Rev. D 51  3870(1995).
\bibitem{GK2005} I.F. Ginzburg, M. Krawczyk, Symmetries of two Higgs doublet model and CP violation, Phys. Rev. D 72, 115013(2005).
\bibitem{K} K. Klimenko, Conditions for certain Higgs potentials to be bounded below, Theor. Math. Phys. 62, 58-65 (1985).
\bibitem{N2020} M. Nebot, Bounded masses in two Higgs doublets models, spontaneous CP violation and $\mathbb{Z}_2$ symmetry, Phys. Rev. D 102, 115002 (2020).
\bibitem{NS}S. Nie, M. Sher, Vacuum stability bounds in the two-Higgs doublet model, Phys. Lett. B 449(1-2), 89-92 (1999).
\bibitem{KKO} S. Kanemura, T. Kasai, Y. Okada, Mass bounds of the lightest CP-even Higgs boson in the two-Higgs-doublet model, Phys. Lett. B 471(2-3), 182-190 (1999).
\bibitem{ERS}D. Eriksson, J. Rathsman, O. Stal, 2HDMC-two-Higgs-doublet model calculator, Comput. Phys. Commun. 181(1), 189-205 (2010); Erratum, Comput. Phys. Commun. .181(5),  (2010).
\bibitem{BBP} R.A. Battye, G.D.   Brawn, A. Pilaftsis, Vacuum topology of the two Higgs doublet model. J. High Energ. Phys. 2011, 20 (2011).
\bibitem{K2016} K. Kannike, Vacuum stability of a general scalar potential of a few fields. Eur. Phys. J. C, 76, 324(2016); Erratum, Eur. Phys. J. C, 78, 355(2018).
\bibitem{K2012}
K. Kannike, Vacuum stability conditions from copositivity criteria. Eur. Phys. J. C, 72, 2093 (2012).
\bibitem{K2021}K. Kannike, Vacuum stability conditions and potential minima for a matrix representation in lightcone orbit space. Eur. Phys. J. C 81, 940 (2021).
\bibitem{GC}G. Chauhan, Vacuum stability and symmetry breaking in left-right symmetric model. J. High Energ. Phys. 2019, 137 (2019).

\bibitem{S2022}Y. Song, Co-positivity of tensors and Stability conditions of CP conserving two-Higgs-doublet potential, Modern Physics Letters A, 38:28n29, 2350130(2023).
\bibitem{BCCIW}H. Bahl, M. Carena, N. M. Coyle, A. Ireland, Carlos E.M. Wagner, New Tools for Dissecting the General 2HDM, J. High Energ. Phys. 2023, 165 (2023).
\bibitem{BFIS} A. Barroso, P.M. Ferreira, I.P. Ivanov, and R. Santos, Metastability bounds on the two Higgs doublet model, J. High Energy Phys. 06,  045(2013).
\bibitem{I2007}I.P. Ivanov, Minkowski space structure of the Higgs potential in the two-Higgs-doublet model,
Phys. Rev. D 75, 035001; Erratum Phys. Rev. D 76, 039902(E) (2007).
\bibitem{GH}J. F. Gunion. H. E. Haber, Conditions for  CP-violation in the general two-Higgs-doublet model,Phys. Rev. D 72, 095002. (2005)
\bibitem{GOO}B. Grzadkowski, O. M. Ogreid, P. Osland, Spontaneous CP-violation in the 2HDM: Physical conditions and the alignment limit, Phys. Rev. D 94 no.11, 115002 (2016). 

\bibitem{MMNN} M. Maniatisa, A. von Manteuffelb, O. Nachtmannc, F. Nagel, Stability and symmetry breaking in the general two-Higgs-doublet model, Eur. Phys. J. C 48, 805-823 (2006).
\bibitem{ACE}  L.E. Andersson,  G. Chang, and T. Elfving, Criteria for copositive matrices using simplices and barycentric coordinates, Linear Algebra Appl. 5, 9-30(1995)
\bibitem{H1983}  K.P. Hadeler, On copositive matrices, Linear Algebra Appl. 49, 79-89(1983).
\bibitem{N1992}  E. Nadler, Nonnegativity of bivariate quadratic functions on a triangle, Comput. Aided Geom. D. 9, 195-205(1992).
\bibitem{LS2022}
J. Liu,  Y. Song, Copositivity for 3rd order symmetric tensors and applications,  Bull. Malays. Math. Sci. Soc. 45(1), 133-152(2022).
\bibitem{SL2022}  Y. Song,  X. Li, Copositivity for a class of fourth order symmetric tensors given by scalar dark matter, J. Optim Theory Appl. 195, 334-346 (2022)
\bibitem{SQ2021}
Y. Song,  L. Qi, Analytical expressions of copositivity for fourth-order symmetric tensors,  Analy. Appl., 19(5), 779-800(2021)
\bibitem{S2021}
Y. Song,  Positive definiteness for 4th order symmetric tensors and applications, Anal. Math. Phys. 11, 10(2021)
\bibitem{Q2005}
L. Qi, Eigenvalues of a real supersymmetric tensor, J. Symbolic Comput., 40(6) 1302-1324(2005)
\bibitem{Q2013}
L. Qi, Symmetric Nonnegative Tensors and Copositive Tensors, Linear Algebra Appl., 439,228-238(2013)
\bibitem{QSZ} L. Qi, Y. Song, X. Zhang, Positivity Conditions for Cubic, Quartic and Quintic Polynomials, J. Nonlinear Convex Anal. 23(2), 191-213(2022)
\bibitem{SQ2015}Y. Song, L. Qi, Necessary and sufficient conditions for copositive tensors, Linear Multilinear A 63(1), 120-131(2015).
\bibitem{QCC2018} L. Qi,  H. Chen,  Y. Chen,  Tensor Eigenvalues and Their Applications, Springer Singapore, 2018.
\bibitem{QL2017} L. Qi,  Z. Luo,  Tensor Analysis: Spectral Theory and Special Tensors, SIAM, Philadelpia 2017.
\bibitem{UW}
G. Ulrich,  L.T. Watson,  Positivity conditions for quartic polynomials. SIAM J. Sci. Comput. 15, 528-544(1994)
	\end{thebibliography}
	\end{document}